\documentstyle[lprocl,11pt]{article}

\def\NPB{{\em Nucl. Phys.} B}
\def\PLB{{\em Phys. Lett.}  B}
\def\PRL{\em Phys. Rev. Lett.}
\def\PRD{{\em Phys. Rev.} D}

\def\be{\begin{equation}}
\def\ee{\end{equation}}
\def\bea{\begin{eqnarray}}
\def\eea{\end{eqnarray}}

\def\del{\partial}                              

\def\frac#1#2{{#1 \over #2}}

\def\half{\ifinner {\scriptstyle {1 \over 2}}
   \else {1 \over 2} \fi}

\def\simge{\mathrel{%
   \rlap{\raise 0.511ex \hbox{$>$}}{\lower 0.511ex \hbox{$\sim$}}}}
\def\simle{\mathrel{
   \rlap{\raise 0.511ex \hbox{$<$}}{\lower 0.511ex \hbox{$\sim$}}}}

\def\slashchar#1{\setbox0=\hbox{$#1$}           
   \dimen0=\wd0                                 
   \setbox1=\hbox{/} \dimen1=\wd1               
   \ifdim\dimen0>\dimen1                        
      \rlap{\hbox to \dimen0{\hfil/\hfil}}      
      #1                                        
   \else                                        
      \rlap{\hbox to \dimen1{\hfil$#1$\hfil}}   
      /                                         
   \fi}                                         %

\def\subrightarrow#1{
  \setbox0=\hbox{
    $\displaystyle\mathop{}
    \limits_{#1}$}
  \dimen0=\wd0
  \advance \dimen0 by .5em
  \mathrel{
    \mathop{\hbox to \dimen0{\rightarrowfill}}
       \limits_{#1}}}                           

\def\journal#1#2#3#4{\ {#1}{\bf #2} ({#3})\  {#4}}

\def\AnnPhys{\journal{{\em Ann.\ Phys.}}}

\def\NPB{\journal{{\em Nucl.\ Phys.\ } {\bf B}}}

\def\PLB{\journal{{\em Phys.\ Lett.\ }{\bf B}}}

\def\PRD{\journal{{\em Phys.\ Rev.\ }{\bf D}}}

\def\PRL{\journal{{\em Phys.\ Rev.\  Lett.}}}

\newcommand \beq{\begin{eqnarray}}
\newcommand \eeq{\end{eqnarray}}
\input epsf
\begin{document}

\begin{titlepage}
\begin{flushright} {Saclay-T96/098}
\end{flushright}
\vspace*{1.5cm}
\begin{center}
\baselineskip=13pt
{\bf THE QUASIPARTICLE STRUCTURE OF HOT GAUGE THEORIES\\}
\vskip0.5cm
Edmond Iancu\\
{\it Service de Physique Th\'eorique\footnote{Laboratoire de la Direction
des
Sciences de la Mati\`ere du Commissariat \`a l'Energie
Atomique}, CE-Saclay \\ 91191 Gif-sur-Yvette, France}\\
\end{center}
\vskip 2.cm
\begin{abstract}
The study of the ultrarelativistic plasmas in perturbation theory
 is plagued with infrared
divergences which are not eliminated by the screening corrections.
They affect, in particular, the computation of the lifetime
of the elementary excitations, thus casting doubt on the
validity of the quasiparticle picture.
We show that, for Abelian plasmas at least,
the infrared problem of the damping rate can be solved 
by a non-perturbative treatment based on the
Bloch-Nordsieck approximation. The resulting expression of the fermion
 propagator is free of divergences, and exhibits
a {\it non-exponential} damping at large times:
$S_R(t)\sim
\exp\{-\alpha T t \ln\omega_pt\}$, where $\omega_p=gT/3$ 
is the plasma frequency and $\alpha=g^2/4\pi$.
\end{abstract}

\vspace*{7.cm}

\begin{flushleft}
Invited talk at the Workshop
 on Quantum Chromodynamics, 3-8 June, 1996,
American University of Paris, France.
\end{flushleft}

\end{titlepage}

\title{THE QUASIPARTICLE STRUCTURE OF HOT GAUGE THEORIES}

\author{Edmond IANCU}

\address{Service de Physique Th\'eorique
\footnote{Laboratoire de la Direction des
Sciences de la Mati\`ere du Commissariat \`a l'Energie Atomique}
, CE-Saclay, \\ 91191 Gif-sur-Yvette, France}

\maketitle\abstracts{
The study of the ultrarelativistic plasmas in perturbation theory
 is plagued with infrared
divergences which are not eliminated by the screening corrections.
They affect, in particular, the computation of the lifetime
of the elementary excitations, thus casting doubt on the
validity of the quasiparticle picture.
We show that, for Abelian plasmas at least,
the infrared problem of the damping rate can be solved 
by a non-perturbative treatment based on the
Bloch-Nordsieck approximation. The resulting expression of the fermion
 propagator is free of divergences, and exhibits
a {\it non-exponential} damping at large times:
$S_R(t)\sim
\exp\{-\alpha T t \ln\omega_pt\}$, where $\omega_p=gT/3$ 
is the plasma frequency and $\alpha=g^2/4\pi$.
}

\section{Introduction}

The study of the elementary excitations of ultrarelativistic  plasmas,
such as  the quark-gluon plasma,  has received much
attention in the recent past \cite{BIO96,MLB96}.
The physical picture which emerges, for both Abelian and non-Abelian
gauge theories, is that of a system with
 two types of degrees of freedom:
{\it i}) the plasma quasiparticles,
whose energy is of the order of the temperature $T$;
{\it ii}) the collective excitations, whose typical energy
is $gT$, where $g$ is the gauge coupling,
assumed to be small: $g\ll 1$ (in QED, $g=e$ is the electric charge).

For this picture to make sense, however, 
it is important that the
 lifetime of the excitations be large compared to the
typical period of the modes. 
Information about the lifetime is obtained from the
retarded propagator. A usual expectation is that
 $S_R(t,{\bf p})$ decays {\it exponentially} in time,
 $S_R(t,{\bf p})\,\sim\,{\rm e}^{-i E(p)t} {\rm e}^{ -\gamma({p}) t}$,
where $E(p) \sim T$ or $gT$ is the average energy of the excitation,
 and $\gamma(p)$ is the damping rate.
The exponential decay may then be associated to a pole
of the Fourier transform  $S_R(\omega,{\bf p})$,
located at $\omega = E-i\gamma\,$:
\beq\label{SRE0}
S_R(\omega, {\bf p})\,=\,
\int_{-\infty}^{\infty} {\rm d}t \,{\rm e}^{-i\omega t}
S_R(t,{\bf p})\,\,\sim\,\, \frac{1}{\omega - (E(p)-i\gamma(p))}\,.\eeq
The quasiparticles are well defined if the damping rate
$\gamma$  is small compared to the energy $E$. If this is the case,
 the respective damping rates
can be computed from the imaginary part of the on-shell
self-energy, $\Sigma(\omega=E(p), {\bf p})$. 

Previous calculations \cite{Pisarski89} suggest that 
$\gamma\sim g^2T$
for both the single-particle and the collective excitations.
In the weak coupling regime $g\ll 1$,
 this is indeed small compared to the corresponding
energies (of order $T$ and $gT$, respectively),
suggesting that the  quasiparticles are well defined, and the
collective modes are weakly damped. However, the computation of 
$\gamma$ in perturbation theory 
is plagued with infrared (IR) divergences, which casts doubt on the
validity of these ${\rm statements}^{\,3-7}$.

The occurence of IR divergences in the perturbation theory 
for hot gauge theories is generic, and reflects the long-range
character of the gauge interactions. However, it is well known that most
of these divergences are actually removed by screening effects
which are generated by the collective motion of the thermal particles.
In ultrarelativistic plasmas, the screening effects manifest themselves
over typical space-time scales $\sim 1/gT$.
 Their inclusion in the perturbative expansion --- which is achieved in a
 gauge-invariant way by resumming the so called ``hard thermal loops''
 (HTL) of Braaten and Pisarski \cite{Pisarski89} ---  greatly
improve the infrared behavior, and yields
IR-finite results for the transport cross-sections \cite{Baym90,BThoma91},
and also for the damping rates of the excitations 
with zero momentum \cite{Pisarski89,KKM}. 

However,  the HTL resummation is not sufficient to render finite
the damping rates of the excitations with non vanishing momenta \cite{Pisarski89}.
The remaining infrared divergences are due to collisions involving the
exchange of longwavelength, quasistatic, magnetic photons (or gluons),
which are not screened in the hard thermal loop approximation.
Such divergences affect the computation of the damping rates
of {\it charged} excitations (fermions and gluons),
 in both Abelian and non-Abelian gauge theories. 
Furthermore, the problem appears for both soft ($p \sim gT$) and hard 
 ($p \sim T$) quasiparticles. In QCD this problem is generally
 avoided by the {\it ad-hoc} introduction of an IR cut-off
(``magnetic screening mass'') $\sim g^2T$, which is
expected to appear dynamically from gluon
self-interactions \cite{MLB96}.
In QED, on the other hand, it is known that no magnetic
screening can occur \cite{Fradkin65},
so that the solution of the problem must lie somewhere else.

We have shown recently \cite{prl} that, for Abelian plasmas,
the divergences can be eliminated through a non perturbative treatment,
which involves a soft photon resummation \`a la Bloch-Nordsieck \cite{Fried}.
We have thus obtained the large-time decay of the fermion
propagator, which is {\it not} of the exponential type alluded
to before, but of the more complicated form
$S_R(t)\,\sim\,{\rm e}^{-iE t} \,{\rm exp}\{-\alpha T  t
\ln\omega_pt \}$, where  $\alpha=g^2/4\pi$ and $\omega_p\sim gT$ is the plasma frequency.
Accordingly, the Fourier transform $S_R(\omega)$ is {\it analytic}
in the vicinity of the mass-shell.
In what follows, I will briefly discuss these results, their derivation and
 their consequences.

\section{Perturbation theory for $\gamma$}

Let me first  recall how the infrared problem
occurs in the perturbative calculation of the damping rate $\gamma$.
For simplicity, I consider an Abelian plasma, as described by QED,
and compute the damping rate of a hard electron, with momentum $p\sim T$
and energy $E(p)=p$.

\begin{figure}
\protect \epsfxsize=8.cm{\centerline{\epsfbox{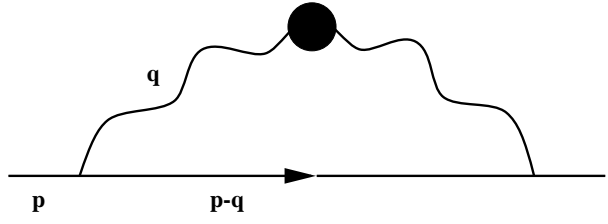}}}
	 \caption{The resummed one-loop self-energy}
\label{effS}
\end{figure}

To leading order in $g$, and after the resummation of the screening 
corrections, $\gamma$ is obtained from the imaginary part of the 
effective one-loop self-energy in Fig.~\ref{effS}.
The blob on the photon line  in this figure
 denotes the  effective  photon propagator in the HTL approximation,
commonly  denoted as ${}^*D_{\mu\nu}(q)$. In the Coulomb gauge, the only non-trivial
components of ${}^*D_{\mu\nu}(q)$
 are the electric (or longitudinal)
one ${}^*D_{00}(q)\equiv {}^*D_l(q)$, and the magnetic (or transverse) one
${}^*D_{ij}(q)=(\delta_{ij}-\hat q_i\hat q_j){}^*D_t(q)$, with
 \beq\label{effd}
{}^*D_l(q_0,q)\,=\,\frac{- 1}{q^2- \Pi_l(q_0,q)},\qquad
{}^*D_t(q_0,q)\,=\,\frac{-1}{q_0^2-q^2 -\Pi_t(q_0,q)},\eeq
where $\Pi_l$ and $\Pi_t$ are the respective pieces
of the photon polarisation tensor \cite{BIO96,MLB96}.
Physically, the on-shell 
discontinuity of the diagram in Fig.~\ref{effS} accounts
for the scattering of the incoming electron off the thermal fermions,
with the exchange of a soft, dressed, virtual photon. 
The corresponding interaction rate is simply computed as
 $\gamma = \sigma \rho$,
where $\rho\sim T^3$ is the density of the scatterers, and $\sigma =\int
{\rm d^2}q \, ({\rm d}\sigma/{\rm d}q^2)$, with $q$ denoting the momentum
of the exchanged (virtual) photon. For a bare (i.e., unscreened)
photon, the Rutherford
formula yields ${\rm d}\sigma/{\rm d}q^2 \sim g^4/q^4$, so that
$\gamma \sim g^4 T^3 \int ({\rm d}q /q^3)$ is quadratically
infrared divergent. Actually, the screening effects included
in  $\Pi_{l,\,t}$ soften this IR behaviour.
We have, in the kinematical regime of interest, ($\omega_p=eT/3$)
\beq\label{pltstatic}
\Pi_l(q_0\ll q) \simeq  3{\omega_p^2}\,\equiv m_D^2,\qquad
\Pi_t(q_0\ll q) \simeq \,-i\,\frac{3\pi}{4}\,{\omega_p^2}\,\frac{q_0}{q}\,
.\eeq 
We see that screening occurs in different ways
in the electric and the magnetic sectors.
In the electric sector, the familiar static Debye screening provides
an IR cut-off $m_{D}\sim gT$. Accordingly,
 the electric contribution to $\gamma$ is finite,
and of the order  $\gamma_l \sim g^4 T^3/m_{D}^2
\sim g^2 T$. In the magnetic sector,
  screening occurs only for nonzero frequency $q_0$.
This comes from the imaginary part of the polarisation tensor,
and corresponds to the absorbtion of the space-like
 photons ($q_0^2<q^2$) by the hard thermal fermions
(Landau damping \cite{PhysKin}).
This  ``dynamical screening'' is not sufficient to completely
remove the IR divergence of $\gamma_t\,$, which is just reduced to a logarithmic one:
\beq\label{G2LR}
\gamma_t &\sim& {g^4 T^3}\,
\int_{0}^{\infty}{\rm d}q  \int_{-q}^q{\rm d}q_0\,|{}^*D_t(q_0,q)|^2\nonumber\\
 &\sim& {g^4 T^3}\,\int_{0}^{\infty}{\rm d}q  \int_{-q}^q{\rm d}q_0
\,\frac{1}{q^4 + (3\pi \omega_p^2 q_0/4q)^2} \,\sim\,
{g^2T}\int_{0}^{\omega_p}\frac{{\rm d}q}{q}\,.\eeq
With an IR cut-off $\mu$, $\gamma_t \sim g^2 T \ln (\omega_p/\mu)$.
The remaining logarithmic divergence is due to collisions involving the
exchange of very soft, {\it quasistatic} ($q_0\to 0$),  magnetic photons,
which are not screened by plasma effects.
To see that, note that the IR contribution to
 $\gamma_t$ comes from momenta $q\ll gT$,
where $|{}^*D_t(q_0,q)|^2$ is almost a delta function of $q_0$:
\beq \label{singDT}
|{}^*D_t(q_0,q)|^2\,\simeq\,
\frac{1} {q^4 + (3\pi \omega_p^2 q_0/4q)^2}\,
\longrightarrow_{q\to 0}\,\frac{4}{3 q \omega_p^2}\,\delta(q_0)\,.\eeq
This is so because,
as $q_0\to 0$, the imaginary part of the polarisation
tensor vanishes {\it linearly}
(see the second equation (\ref{pltstatic})), 
a property which can be related to the behaviour of the
phase space for the Landau damping processes.
Since energy conservation requires $q_0=q\cos\theta$, 
where $\theta$ is the angle
between the momentum of the virtual photon (${\bf q}$) and that
of the incoming fermion (${\bf p}$), 
the magnetic photons which are responsible for the singularity
are emitted, or absorbed, at nearly 90 degrees.

\section{A non perturbative calculation}

\begin{figure}
\protect \epsfxsize=14.cm{\centerline{\epsfbox{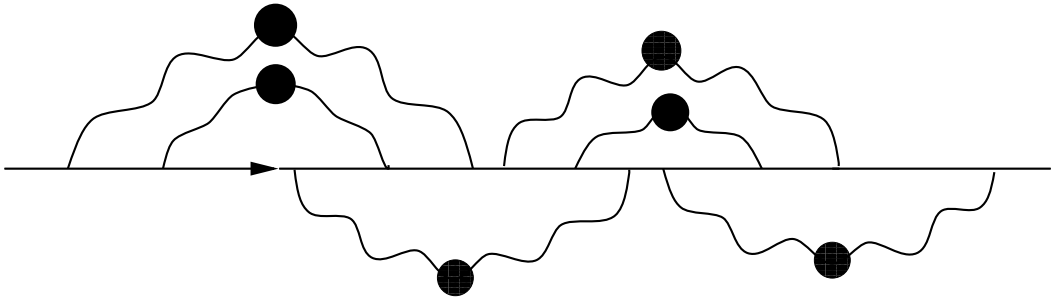}}}
	 \caption{A generic $n$-loop diagram (here, $n=6$)
for the self-energy in quenched QED.}
\label{effN}
\end{figure}

The IR divergence of the leading order calculation
invites to a more thorough investigation
of the higher orders contributions to $\gamma$. Such an
analysis \cite{prl,debye} reveals strong, power-like, infrared
divergences, which signal the breakdown of the perturbation theory.
To a given order in the loop expansion, the most
singular contributions to $\gamma$ arise from the quenched
(no internal fermion loops) self-energy diagrams
in Fig.~\ref{effN}, where all the internal photon lines are of the magnetic type.
As in the one loop calculation, the {\it leading} divergences arise,
in all orders, from the kinematical regime where the
internal photons are soft ($q\to 0$) and quasistatic ($q_0\to 0$).
Physically, these divergences come from multiple magnetic collisions.

This peculiar kinematical regime can be conveniently
exploited in the imaginary time formalism \cite{MLB96}, 
where the internal photon lines carry only discrete
(and purely imaginary) energies, of the form $q_0=i \omega_n=
i 2\pi n T$, with integer $n$ (the so-called  Matsubara frequencies).
Note that the non-static modes, with $n\ne 0$, are well separated from
the static one $q_0=0$ by a gap of order $T$. 
In this formalism, all the leading IR divergences of the damping rate --- which,
I recall, arise from the kinematical limit $q_0\to 0$ ---
 are concentrated in diagrams in which
the photon lines are static, i.e., they 
 carry zero Matsubara frequency \cite{Marini92,prl}.
Note that, for these diagrams, all the loop integrations 
are three-dimensional (they run over
the three-momenta of the internal photons), so that the associated IR divergences
are those of a three-dimensional gauge theory. This clearly
emphasizes the non perturbative character of the leading IR structure.

In what follows, we restrict
ourselves to these diagrams, and compute their contribution
to the fermion propagator near the mass-shell, in a non perturbative way.
This can be ``exactly'' done in the Bloch-Nordsieck approximation \cite{Fried},
which is the relevant approximation for the infrared structure
of interest. Namely, since the incoming fermion is interacting
only with very soft ($q\to 0$) static ($q_0=0$) magnetic photons,
its trajectory is not significantly deviated by the
successive collisions, and its spin state does not change.
Thus, we can ignore the spin degrees of freedom,
which play no dynamical role, and we can assume the fermion to
move along a straightline trajectory with constant velocity
${\bf v}$ (for the ultrarelativistic hard  fermion,
$|{\bf v}|=1$; more generally, for the soft excitations,
${\bf v}(p) \equiv \del E(p)/\del {\bf p} = v(p) {\bf \hat p}$
 is the corresponding group velocity, with $|v(p)|< 1$).
Under these assumptions, the fermion propagator can be easily
computed as \cite{prl}
\beq\label{SRT}
S_R(t,{\bf p})&=&i\,\theta(t) {\rm e}^{-iE(p)t}\,\Delta(t),\eeq
where
\beq\label{SR0}
\Delta(t)=  {\rm exp}\left \{-g^2T
\int^{\omega_p} \frac{{\rm d}^3q}{(2\pi)^3} 
\,\frac{1}{q^2}\,\frac
{1-  {\rm cos}\,t ({\bf v}(p) \cdot {\bf q})}{
({\bf \hat p \cdot q})^2} \right\},\eeq
contains all the non-trivial time dependence. 
The integral in eq.~(\ref{SR0}) is formally identical to that one would 
get in the Bloch-Nordsieck model in 3 dimensions.
Note, however, the upper cut-off $\omega_p \sim gT$, which occurs for
the same reasons as in eq.~(\ref{G2LR}). Namely, it reflects the dynamical
cut-off at momenta $\sim gT$, as provided by the Landau damping.

The integral over $q$ has no infrared divergence, but one can verify 
that the expansion of $\Delta(t)$ in powers of $g^2$ generates 
the most singular pieces of the usual perturbative expansion
for the self-energy \cite{prl}.
Because our approximations preserve only
 the leading infrared behavior of the perturbation theory,
eq.~(\ref{SR0}) describes  only the leading {\it large-time} ($t\gg 1/gT$) behavior
of $\Delta(t)$. This is gauge independent  \cite{prl} and of the form (we set here $\alpha
= g^2/4\pi$ and  $v(p)=1$ to simplify writing)
\beq\label{DLT}
\Delta(\omega_pt\gg 1)\,\simeq \,{\rm exp}\Bigl( -\alpha Tt \ln \omega_p
t\Bigr).\eeq
Thus, contrary to what perturbation theory predicts, 
$\Delta(t)$ is decreasing faster than any exponential. It follows that
 the  Fourier transform 
\beq\label{SRE}
S_R(\omega, {\bf p})\,=\,
\int_{-\infty}^{\infty} {\rm d}t \,{\rm e}^{-i\omega t}
S_R(t,{\bf p})\,=\,
i\int_0^{\infty}{\rm d}t
\,{\rm e}^{it(\omega- E(p)+i\eta)}\,\Delta(t),\eeq
  exists 
for {\it any} complex (and finite) $\omega$. Thus,  the retarded propagator
 $S_R(\omega)$ is an entire function, with sole singularity at 
Im$\,\omega\to -\infty$.   The associated spectral density
$\rho(\omega, p)$ (proportional to the imaginary part of
$S_R(\omega, {\bf p})$) retains the shape
of a {\it resonance}  strongly peaked around the perturbative mass-shell 
$\omega = E(p)$, 
with a typical width of order $\sim g^2T \ln(1/g)$ \cite{prl}.

\section{Conclusions}

The previous analysis solves the IR problem of the damping rate,
thus confirming the quasiparticle picture for the hot Abelian plasmas.
For high temperature QCD, on the other hand, the resolution of the
corresponding problem
also requires the understanding of the non-perturbative sector of the
magnetostatic gluons.

As a crude model of QCD, let us assume, in agreement
with heuristic arguments  \cite{MLB96}, and also with  lattice
computations \cite{Tar}, that a screening mass $\mu\sim g^2T$
is dynamically generated in the magnetic sector.
After replacing $1/q^2 \to
1/(q^2 + \mu^2)$ for the photon propagator in eq.~(\ref{SR0}),
this equation provides,
{\it at very large times} $t\simge 1/g^2T$,
an exponential decay:  $\Delta(t)
\sim \exp(-\gamma t)$, with $\gamma = \alpha T\ln(\omega_p/\mu)
=  \alpha T\ln(1/g)$.
However, in the physically more interesting regime of {\it intermediate
times}  $1/gT \ll t \ll 1/g^2 T$, the behavior is non-exponential and 
governed by the plasma frequency, according to our result
(\ref{DLT}): $\Delta(t)\sim \exp ( -\alpha Tt \ln \omega_p
t)$.  Thus, at least within this limited model,
which is QED with a ``magnetic mass'', the time behavior
in the physical regime remains controlled by the
Bloch-Nordsieck mechanism. But, of course, this result gives no
serious indication about the real situation in QCD, since
it is unknown whether, in the present problem,
 the effects of the gluon self-interactions
can be simply summarized in terms of a magnetic mass.


\section*{References}

\begin{thebibliography}{9}

\bibitem{BIO96}
 J.P. Blaizot, J.-Y. Ollitrault and E.~Iancu, {\em Collective
Phenomena in the Quark-Gluon Plasma}, in {\em Quark-Gluon Plasma 2},
ed. R.C. Hwa  (World Scientific, Singapore, 1996).

\bibitem{MLB96}
M. Le Bellac, {\it Recent Developments in Finite Temperature Quantum
 Field Theories}
(Cambridge University Press, 1996).

\bibitem{Pisarski89}
R.D.~Pisarski, \PRL{63}{1989}{1129};
E.~Braaten and R.D.~Pisarski, \PRL{64}{1990}{1338}; 
 \PRD{42}{1990}{2156}; \NPB{337}{1990}{569}.
 
\bibitem{Lebedev}
V.V Lebedev and A.V. Smilga, \PLB{253}{1991}{231}; \AnnPhys{202}{1990}{229};
{\em Physica} {\bf A181} (1992) 187.

\bibitem{Marini92}
C.P. Burgess and A.L. Marini,
 \PRD{45}{1992}{R17}; A.K. Rebhan, \PRD{46}{1992}{482};
F. Flechsig, H. Schulz and A.K. Rebhan,
 \PRD{52}{1995}{2994}.


\bibitem{Altherr93}
R.D. Pisarski, \PRD{47}{1993}{5589};
T. Altherr, E. Petitgirard and T. del Rio Gaztelurrutia, 
\PRD{47}{1993}{703}; H.~Heiselberg and C.J.~Pethick, \PRD{47}{1993}{R769};
A. Ni\'egawa, \PRL{73}{1994}{2023};
K. Takashiba, hep-ph/9501223 (unpublished).


\bibitem{Pilon93}
S. Peign\'e, E. Pilon and D. Schiff, 
{\em Z. Phys.} {\bf C60} (1993) 455;
A.V. Smilga, {\em Phys. Atom. Nuclei} {\bf 57} (1994) 519;
R. Baier and R. Kobes, \PRD{50}{1994}{5944}.


\bibitem{Baym90}
G.~Baym, H.~Monien, C.J.~Pethick, and D.G.~Ravenhall,\PRL{64}{1990}{1867}.

\bibitem{BThoma91}
E.~Braaten and M.H.~Thoma, \PRD{44}{1991}{1298}.

\bibitem{KKM}
R.~Kobes, G.~Kunstatter and K.~Mak, \PRD{45}{1992}{4632}; E.~Braaten and
R.D.~Pisarski, \PRD{46}{1992}{1829}.

\bibitem{Fradkin65}
E. Fradkin, {\em Proc. Lebedev Phys. Inst.} {\bf 29} (1965) 7;
J.P. Blaizot, E. Iancu and R. Parwani, \PRD{52}{1995}{2543}.


\bibitem{prl}
J.P. Blaizot and E. Iancu, \PRL{76}{1996}{3080}
and hep-ph/9607303, to appear in {\em Phys. Rev.} {\bf D}.

\bibitem{Fried}
H.M. Fried, {\em Functional Methods and Models in Quantum
Field Theory} (the MIT Press, Boston, 1972).

\bibitem{PhysKin}
E.M.~Lifshitz and L.P.~Pitaevskii,  {\em Physical Kinetics} (Pergamon Press,
Oxford, 1981).

\bibitem{debye}
J.P. Blaizot and E. Iancu, \NPB{459}{1996}{559}.


\bibitem{Tar}
C. DeTar, {\em Quark-Gluon Plasma in Numerical Simulations of Lattice QCD},
in  {\em Quark-Gluon Plasma 2}, ed. R.C. Hwa (World Scientific, Singapore, 1996).

\end{thebibliography}
\end{document}